\documentclass[preprint,review,12pt]{revtex4}
\usepackage{amsmath}
\usepackage[dvips]{graphicx}
\begin{document}
\title{Effects of Bridge Functions on Radial Distribution Functions of Liquid Water}

\author{Shigenori TANAKA}
\email{tanaka2@kobe-u.ac.jp}
\affiliation{Graduate School of System Informatics, Kobe University, 1-1 Rokkodai, Nada-ku, Kobe 657-8501, Japan}

\author{Miki NAKANO}
\affiliation{Frontier Institute for Biomolecular Engineering Research (FIBER), Konan University, \\ 
7-1-20 Minatojima-minamimachi, Chuo-ku, Kobe 650-0047, Japan}

\date{\today}

\begin{abstract}
In this report 
the radial distribution functions (RDFs) of liquid water are calculated 
on the basis of the classical density functional theory 
combined with the reference interaction site model for molecular liquids.
The bridge functions, which are neglected in the hypernetted-chain (HNC) approximation, are taken into account through 
the density expansion for the Helmholtz free energy functional up to the third order. 
A factorization approximation to 
the ternary direct correlation functions in terms of the 
site-site pair correlation functions is then employed 
in the expression of the bridge functions, thus leading to a closed set of integral equations for the determination of the RDFs. 
It is confirmed through numerical calculations that incorporation of the oxygen-oxygen bridge function substantially improves 
the poor descriptions by the HNC approximation at room temperature, 
{\it e.g.}, for the second peak of the oxygen-oxygen RDF. \\


\end{abstract}

\maketitle

\setlength{\baselineskip}{24pt}

\section{Introduction}

Integral equation approach provides an efficient tool to calculate the correlation and thermodynamic functions of liquids 
with high accuracy and mild cost of computation (Hansen and McDonald, 2006; Ichimaru {\it et al.}, 1987).
This analytical approach often gives fair results on par with those by extensive Monte Carlo or molecular dynamics (MD) simulations, 
and combined with the reference interaction site model (RISM) method (Chandler and Andersen, 1972; Hansen and McDonald, 2006; Hirata and Rossky, 1981), 
can comprehensively describe the equilibrium properties of molecular liquids such as liquid water (Pettitt and Rossky, 1982), 
which plays essential roles in a variety of biochemical processes. 
Recent developments in the methods, algorithms and benchmarks 
(Lombardero {\it et al.}, 1999; Lue and Blankschtein, 1995; Reddy {\it et al.}, 2003; Richardi {\it et al.}, 1999; Sato, 2013; Sumi and Sekino, 2006) 
indicate that the RISM-based integral equation approach can provide an alternative route for 
theoretical analyses on water and related aqueous systems with comparable reliability to more expensive, computer simulation approaches.
However, it has also been observed 
(Lombardero {\it et al.}, 1999; Lue and Blankschtein, 1995; Reddy {\it et al.}, 2003; Richardi {\it et al.}, 1999; Sato, 2013; Sumi and Sekino, 2006) 
that the descriptions of intermolecular correlations of water 
become less accurate at room temperature in comparison with at higher temperatures.

\par

Among the RISM-based integral equation formalisms for molecular liquids, the density functional theory (DFT) approach 
due to Chandler, McCoy and Singer (Chandler {\it et al.}, 1986) is one of the most sophisticated and dependable methods.
Donley, Curro and McCoy (DCM) (Donley {\it et al.}, 1994) then extended this DFT scheme for the calculations of pair correlation functions of molecular liquids, and later, 
Reddy {\it et al.} (Reddy {\it et al.}, 2003) and Sumi {\it et al.} (Sumi and Sekino, 2006) applied this method to the calculations of the radial distribution functions 
(RDFs) of fluid water.
Their calculated results demonstrated that the DCM scheme provides accurate descriptions at high temperatures, but the deviations from the computer simulation results 
are observed at room temperature, {\it e.g.}, in the description of the location of the second peak of oxygen-oxygen (O-O) RDF, 
which may be associated with the formation of tetrahedral structure of condensed water.
Considering that the DCM theory is based on the density expansion of free energy functional up to the second order, which corresponds to 
the hypernetted-chain (HNC) approximation (Hansen and McDonald, 2006; Ichimaru {\it et al.}, 1987) in liquid theory, the present authors (Tanaka and Nakano, 2013) have recently extended 
the DCM theory so as to retain the density expansion of free energy functional up to the third order, thus 
taking into account the effect of the bridge functions (Hansen and McDonald, 2006; Ichimaru {\it et al.}, 1987) beyond the HNC approximation.
Their preliminary test calculation in which a simple Gaussian approximation is employed in the factorization (Barrat {\it et al.}, 1988) for the ternary direct correlation functions 
in the bridge functions has, 
however, failed to appropriately reproduce the simulation results for liquid water at room temperature, suggesting the 
importance of incorporation of detailed structures of correlation functions in the representation of ternary direct correlation functions.
The present work then investigates another factorization approximation 
for the bridge functions, in which the site-site pair correlation functions are employed in the 
factorization according to a scheme (Iyetomi and Ichimaru, 1983) that worked well in the one-component plasma system (Ichimaru {\it et al.}, 1987).

\section{Methods}

Let us consider a liquid water system consisting of rigid ${\rm H}_{2}{\rm O}$ molecules whose average number density and temperature are $\rho$ and $T$, respectively.
We then assume the site-site interaction potentials $v_{\alpha\beta}(\mbox{\boldmath $r$})$ between 
the atomic sites $\alpha$ and $\beta$ ($\alpha, \beta = $ O, H, H, {\it i.e.}, one oxygen and two hydrogens) on two water molecules.
According to the DFT scheme (Donley {\it et al.}, 1994; Tanaka and Nakano, 2013), the RDF 
between the sites $\alpha$ and $\beta$ is given by 

\begin{equation}
g_{\alpha\beta}\left(\mbox{\boldmath $r$}_{\alpha}-\mbox{\boldmath $R$}_{\beta}\right) = 
\left\langle\left\langle\exp\left[-\beta\sum_{\eta,\gamma}v_{\eta\gamma}\left(\mbox{\boldmath $r$}_{\eta}-\mbox{\boldmath $R$}_{\gamma}\right)
+ \Lambda\left(\left\{\mbox{\boldmath $r$}\right\},\left\{\mbox{\boldmath $R$}\right\}\right) 
+ B\left(\left\{\mbox{\boldmath $r$}\right\},\left\{\mbox{\boldmath $R$}\right\}\right)\right]
\right\rangle\right\rangle_{\mbox{\boldmath $r$}_{\alpha},\mbox{\boldmath $R$}_{\beta}}^{\mathcal{P}}
\end{equation}

\noindent
with $\beta = 1/k_{\rm{B}}T$ and $k_{\rm{B}}$ being the Boltzmann constant, 
where $\left\langle\left\langle \enskip 
\right\rangle\right\rangle_{\mbox{\boldmath $r$}_{\alpha},\mbox{\boldmath $R$}_{\beta}}^{\mathcal{P}}$ 
refers to the avarage over the relative orientation of two water molecules whose $\alpha$th and $\beta$th sites 
are fixed at $\mbox{\boldmath $r$}_{\alpha}$ and $\mbox{\boldmath $R$}_{\beta}$, respectively, 
and whose coordinates are represented by $\{\mbox{\boldmath $r$}\}=\{\mbox{\boldmath $r$}_{\eta}\}$ 
and $\{\mbox{\boldmath $R$}\}=\{\mbox{\boldmath $R$}_{\gamma}\}$, respectively. 
In this expression, 
the two-body part is given by 

\begin{equation}
\Lambda\left(\left\{\mbox{\boldmath $r$}\right\},\left\{\mbox{\boldmath $R$}\right\}\right) 
= \rho\sum_{\eta,\gamma}\sum_{\sigma,\xi}\int d\mbox{\boldmath $r$}_{\sigma}\int d\mbox{\boldmath $r$}_{\xi}
c_{\eta\sigma}(\mbox{\boldmath $r$}_{\eta}-\mbox{\boldmath $r$}_{\sigma})
h_{\sigma\xi}(\mbox{\boldmath $r$}_{\sigma}-\mbox{\boldmath $r$}_{\xi})
\omega_{\xi\gamma}^{-1}(\mbox{\boldmath $r$}_{\xi}-\mbox{\boldmath $R$}_{\gamma}) 
\end{equation}

\noindent
with the aid of a site-site pairwise approximation for the induced density (Donley {\it et al.}, 1994), where 

\begin{equation}
\omega_{\alpha\beta}(\mbox{\boldmath $r$}) = \delta_{\alpha\beta}\delta(\mbox{\boldmath $r$}) + 
(1-\delta_{\alpha\beta})\delta(r-L_{\alpha\beta})/4\pi L_{\alpha\beta}^{2}
\end{equation}

\noindent
with $L_{\alpha\beta}$ being the distance between sites $\alpha$ and $\beta$ means the intramolecular correlation functions.
The site-site pair correlation functions $h_{\alpha\beta}(r) = g_{\alpha\beta}(r) - 1$ and the two-body direct correlation functions 
$c_{\alpha\beta}(r)$ are then related via the site-site Ornstein-Zernike or RISM equation (Chandler and Andersen, 1972; Hansen and McDonald, 2006), 

\begin{equation}
h_{\alpha\beta}(r) = \sum_{\eta,\sigma}\left[\omega_{\alpha\eta}*c_{\eta\sigma}*\omega_{\sigma\beta}(r) 
+ \rho\omega_{\alpha\eta}*c_{\eta\sigma}*h_{\sigma\beta}(r)\right],
\end{equation}

\noindent
where $*$ means the convolution. 
If we neglect the higher-order contribution $B\left(\left\{\mbox{\boldmath $r$}\right\},\left\{\mbox{\boldmath $R$}\right\}\right)$ in Eq.\ (1), 
we are led to a two-body approximation that is equivalent to the HNC approximation (Hansen and McDonald, 2006; Ichimaru {\it et al.}, 1987).
The HNC solutions to the integral equations for the correlation functions of fluid water 
were obtained and compared to the computer simulation and experimental results in the literature 
(Reddy {\it et al.}, 2003; Sumi and Sekino, 2006; Tanaka and Nakano, 2013).
In contrast to successful descriptions at higher temperatures, one then observed a significant discrepancy between the HNC and MD results at room temperature, 
in particular concerning the location of the second peak of $g_{OO}(r)$ that was situated at $r \simeq 5.5$ \AA\  in the HNC result in comparison with 
$r \simeq 4.5$ \AA\  in the MD and experimental results.
This observation indicates that the O-O tetrahedral ordering in dense, low-temperature water systems is not appropriately described 
by the second-order, HNC approximation for the intermolecular correlations.

\par

Then, we proceed to an incorporation of the triplet correlation term in Eq.\ (1). 
In a preceding DFT study (Tanaka and Nakano, 2013) we derived an explicit expression for 
$B\left(\left\{\mbox{\boldmath $r$}\right\},\left\{\mbox{\boldmath $R$}\right\}\right)$ 
in terms of ternary direct correlation functions, and employed,   
as a first attempt, a simple Gaussian approximation for the factorized, three-body direct correlation functions. 
However, the calculated result for $g_{\alpha\beta}(r)$ failed to quantitatively reproduce the MD or experimental results. 
In the present study we instead consider another factorization approximation according to Iyetomi and Ichimaru (Iyetomi and Ichimaru, 1983), 
which took relevant account of the lowest-order contribution and worked well in the case of strongly coupled, one-component plasma (Ichimaru {\it et al.}, 1987).
It is expected that this approximation would provide a good description for hard and soft sphere systems as well as for Coulombic system, 
considering the universality in strongly correlated systems (Hansen and McDonald, 2006; Ichimaru {\it et al.}, 1987).
Taking into account only the O-O contribution, which would play a primary role, the bridge function can then be given by 

\begin{eqnarray}
B_{OO}(r) 
= \frac{1}{2}\rho^{2}\int d\mbox{\boldmath $r$}'\int d\mbox{\boldmath $r$}''
\ h_{OO}(\mbox{\boldmath $r$}-\mbox{\boldmath $r$}')
h_{OO}(\mbox{\boldmath $r$}'-\mbox{\boldmath $r$}'')h_{OO}(\mbox{\boldmath $r$}''-\mbox{\boldmath $r$})h_{OO}(r')h_{OO}(r'') 
\end{eqnarray}

\noindent
with $r = \vert\mbox{\boldmath $r$}\vert = \vert\mbox{\boldmath $r$}_{O} - \mbox{\boldmath $R$}_{O}\vert$ 
for $B\left(\left\{\mbox{\boldmath $r$}\right\},\left\{\mbox{\boldmath $R$}\right\}\right)$ in Eq.\ (1), 
which is symmetric with respect to $\mbox{\boldmath $r$}_{O}$ and $\mbox{\boldmath $R$}_{O}$ due to the exclusive consideration of O-O correlation. 

\par

However, there is a drawback in the expression of Eq.\ (5) that this form of bridge function does not reproduce the correct short-range bahavior 
due to $h_{\alpha\beta}(r) \to -1$ at $r \to 0$ (Barrat {\it et al.}, 1988; Iyetomi and Ichimaru, 1983). 
We hence renormalize the bridge function as (Tanaka and Nakano, 2013) 

\begin{equation}
\tilde{B}(r) = f(r)B_{OO}(r)
\end{equation}

\noindent
by introducing a short-range enhancement function (Iyetomi and Ichimaru, 1983) as 

\begin{equation}
f(r) = (C-1)\exp\left[-\left(\frac{r}{a}\right)^{2}\right]+1. 
\end{equation}

\noindent
The parameters in Eq.\ (7) are roughly fixed as $a = 4.5$ \AA \ and $C = 30.0$ in the following analysis, 
considering the location of the second peak of $g_{OO}(r)$ for the former and the typical magnitude of the short-range renormalization for $h_{\alpha\beta}(r)$ 
(Barrat {\it et al.}, 1988; Iyetomi and Ichimaru, 1983) for the latter.
In addition, to reduce the computational cost, we use the 
experimental results (Soper, 2000) for $h_{OO}(r)$ to evaluate $B_{OO}(r)$ in terms of Eq.\ (5).
The behavior of $\tilde{B}(r)$ at $\rho = 0.0334$ \AA$^{-3}$ and $T = 300$ K is depicted in Fig.\ 1.

\section{Results and discussion} 

We have employed the SPC/E model (Berendsen {\it et al.}, 1987) for the (rigid) structure and intermolecular potentials for water molecules, 
thus assuming a three-site (one oxygen and two hydrogens) model in which the Lennard-Jones potential works only between the oxygen (O) sites.
We have thus carried out the calculation of the RDFs of liquid water at $\rho = 0.0334$ \AA$^{-3}$ and $T = 300$ K, 
where the renormalized bridge function, $\tilde{B}(r)$, above was employed in Eq.\ (1).
The computational details are found in the literature (Tanaka and Nakano, 2013).
The calculated results for $g_{\alpha\beta}(r)$ are shown in Fig.\ 2, in which 
we have also depicted the results by 10 ns MD simulation (Tanaka and Nakano, 2013) for the SPC/E water performed with {\it AMBER 12} software (Case {\it et al.}, 2012) 
as well as those by a neutron diffraction experiment (Soper, 2000) for comparison.

\par

It is seen in Fig.\ 2 (b) and (c) that the present DFT scheme with inclusion of $\tilde{B}(r)$ 
reproduces the behaviors of $g_{OH}(r)$ and $g_{HH}(r)$ obtained in the MD simulation 
fairly well. 
Concerning the description of $g_{OO}(r)$ seen in Fig.\ 2 (a), on the other hand, the present DFT scheme 
slightly improves the behavior of the second peak of $g_{OO}(r)$ over the HNC result (Reddy {\it et al.}, 2003; Sumi and Sekino, 2006; Tanaka and Nakano, 2013), 
yielding the (weak) peak at $r \simeq 4.7$ \AA\  in comparison with that at $r \simeq 4.5$ \AA\  in the MD result 
and $r \simeq 5.5$ \AA\  in the HNC result. 
It should be remarked, however, that the agreement between the DFT and MD results is still unsatisfactory 
especially concerning the quantitative description of the oscillatory behavior of $g_{OO}(r)$, 
{\it e.g.}, the location of the third peak, 
thus needing further improvement on 
the theoretical scheme 
beyond the factorization approximation in terms of $h_{OO}(r)$.
For instance, the contributions other than the O-O correlation may play some roles in 
$B\left(\left\{\mbox{\boldmath $r$}\right\},\left\{\mbox{\boldmath $R$}\right\}\right)$. 
In addition, there is a possibility that the fourth- and higher-order terms contribute to the bridge functions to some extent.

\par

In this context, we have attempted to correct $\tilde{B}(r)$ in an {\it ad hoc} way by 

\begin{equation}
\delta\tilde{B}(r) = D\left\{\exp\left[-\left(\frac{r-r_{1}}{\Delta_{1}}\right)^{2}\right] - \exp\left[-\left(\frac{r-r_{2}}{\Delta_{2}}\right)^{2}\right] 
- \exp\left[-\left(\frac{r-r_{3}}{\Delta_{3}}\right)^{2}\right]\right\} 
\end{equation}

\noindent
with $D=0.1$, $r_{1}=4.5$ \AA, $r_{2}=3.5$ \AA, $r_{3}=5.5$ \AA, and $\Delta_{1}=\Delta_{2}=\Delta_{3}=0.5$ \AA, 
considering the discrepancy between the DFT and MD results for $g_{OO}(r)$ observed in Fig.\ 2. 
We have thus performed the calculation of RDFs with the corrected bridge function, 

\begin{equation}
\tilde{B}^{c}(r) = \tilde{B}(r) + \delta\tilde{B}(r).
\end{equation}

\noindent
While the correction $\delta\tilde{B}(r)$ is very minor as seen in Fig.\ 1 (a) (indiscernible between $\tilde{B}(r)$ and $\tilde{B}^{c}(r)$ at the full scale), 
we have found that this correction substantially 
improves the agreement between the DFT and MD results as observed in Fig.\ 2, 
which indicates a delicate manner (see Fig.\ 1 (b) at the zoomed scale) to quantitatively describe the RDFs with the bridge functions at medium to long distances. 
More accurate descriptions of the RDFs for liquid water would thus be a challenging task in the framework of many-body theory.

\section{Conclusion}

In summary, we calculated the RDFs of liquid water at ambient temperature and density on the basis of the third-order 
DFT method, in which the ternary direct correlation functions were expressed by the factorization with the pair correlation functions 
and the O-O bridge function was taken into account beyond the HNC approximation in the expression for the RDFs. 
Improvement on the HNC scheme was thus achieved with the use of the proposed bridge function and its corrected one, 
especially concerning the behavior of $g_{OO}(r)$, while some future tasks still remain for better theoretical description.

\section*{Acknowledgment}

\noindent
S.T. would like to acknowledge the Grants-in-Aid for Scientific Research (No.\ 23540451) from the Ministry of Education, Cultute, Sports, Science and Technology (MEXT).

\newpage


\section*{Figure caption}

\noindent
Fig.\ 1. Oxygen-oxygen bridge functions $\tilde{B}(r)$ (red dashed line) and $\tilde{B}^{c}(r)$ (blue solid line) for liquid water at $\rho = 0.0334$ \AA$^{-3}$ and 
$T = 300$ K. (a) Full scale. (b) Zoomed scale.

\vspace{0.5cm}

\noindent
Fig.\ 2. Radial distribution functions (RDFs) obtained by the DFT scheme with inclusion of bridge function 
$\tilde{B}(r)$ (red dotted line), that with $\tilde{B}^{c}(r)$ (blue solid line), 
the MD simulation (Tanaka and Nakano, 2013) (green dashed lines), and the neutron diffraction experiment (Soper, 2000) (circles) for liquid water at $\rho = 0.0334$ \AA$^{-3}$ and 
$T = 300$ K. 
(a) $g_{OO}(r)$; (b) $g_{OH}(r)$; (c) $g_{HH}(r)$.

\end{document}